\documentclass[proceedings]{JHEP3} 
\PrHEP{hep2001}
\usepackage{epsfig}                    

\newcommand{\GeV}{\rm GeV}
\newcommand{\an}{\overline{\alpha}_0}
\newcommand{\as}{\alpha_s}
\newcommand{\asmz}{\alpha_s(M_Z)}
\newcommand{\mean}[1]{\langle #1 \rangle}
\newcommand{\qmean}{\mean{Q}}
\def\hc{{\cal H}_{\mathrm{C}}}

\conference{EPS HEP 2001}

\title{Resummed event shape spectra in DIS}

\author{Mrinal Dasgupta$\,^a$, 
        \speaker{Hans-Ulrich Martyn}$\,^b$\\
        $^a$ DESY, Theory Group, Hamburg, Germany \\ 
        $^b$ I. Physikalisches Institut der RWTH,  Aachen, Germany } 
      
\abstract{Resummed results are discussed for DIS event shape variable distributions involving the thrust defined using the thrust axis of the current hemisphere, the $C$ para\-meter and the jet mass. The resummation of these variables requires the development of some techniques different to those applied before for the resummation of DIS event shapes. Including power corrections and matching a comparison with H1 data is presented.}

\begin{document}

\section{Introduction}
The study of event shape variable distributions is a rather well developed 
area within perturbative QCD. It was observed nearly a decade ago 
(see e.g. \cite{CTTW}) that shape variable distributions, although infrared and collinear safe observables suffer from the presence of logarithms that spoil 
the convergence of the fixed order expansion. 
For a typical event shape cross section, which is the integral of the distribution over a limited range of the shape variable $V$, 
 the leading behaviour at $n^{th}$ order in perturbation theory is 
\begin{equation}
\label{dl}
R(V) \sim \alpha_s^n \ln^{2n} \frac{1}{V}+\ldots
\end{equation}
The dots represent less singular terms which, however, are still relevant for a description of the data in the kinematic region $V \ll 1$, the two-jet (1+1 jet ) limit in $e^{+}e^{-}$ annihilation (DIS).
These logarithms can, in many cases, be resummed into an expression that is both meaningful (physically) and is practically useful for comparisons with data.
The state of the art resummation is generally till single logarithmic accuracy, i.e. aims to account for all terms more singular than and upto $\alpha_s^{n} \ln^{n} \frac{1}{V}$.

Resumming the double logarithms (Eq.~\ref{dl}) is straightforward and involves 
the implementation of a probabilistic parton branching evolution pattern on each jet which method also treats collinear enhancements which are sources of single logarithms. 
However, in certain cases 
the resummation of single logarithms can be more involved. Recent examples are provided by three-jet observables in $e^{+}e^{-}$  
annihilation \cite{yuri1} and by a class of observables sensitive to radiation in only a portion of phase space \cite{dassal}. 
The observables we discuss here are of this latter kind.

\section{Definitions}
Next we shall define the various event shapes that have most recently been resummed.
These are the thrust wrt the thrust axis in the Breit frame, the jet mass and the  $C$ parameter, defined respectively as
\begin{equation}
  T = \max_{\vec{n}} \frac {\sum_{\hc}  |\vec{P_i}.\vec{n}|}{\sum_{\hc}  |\vec{P_i}|} \ ,
\end{equation}
\begin{equation}
  \label{eq:rhodef}
  \rho = \frac {\left(\sum_{\hc}  P_i\right)^2}{4 \left(\sum_{\hc}
      |\vec{P}_i|\right)^2 } \ ,
\end{equation}
\begin{equation}
\label{eq:Cparam}
C = \frac{3}{2}  
\frac{\sum_{a,b \hc}|\vec{P_a}||\vec{P_b}| \sin^2 \theta_{ab}}{\left (\sum_{\hc}
      |\vec{P_i}|\right)^2} \ ,
\end{equation}
where $P_i$ are final state hadron momenta and the sums count only particles in
$\hc$, the current hemisphere. 
Although the above definitions are in terms of the hadron momenta, 
perturbative calculations are performed by replacing the hadron momenta  
by parton momenta. This procedure costs us non-perturbative effects 
that vary inversely 
as the hard scale $Q$ of the process (power corrections) 
and which we estimate 
through renormalon based techniques. 
Another definition of the thrust involves using the photon axis in the Breit frame $\vec{n}_\gamma$ and was resummed earlier \cite{ADS}
\begin{equation}
\label{thrnor}  
T_\gamma = 2 \frac{\sum_{\hc}  |\vec{P_i}.\vec{n}_\gamma|}{Q} \ .
\end{equation}

We neglect $Z$ boson exchange although this could be relevant at higher $Q$ values. The resummed results (form factors) will in any case be identical to the photon case 
which means that the dominant logarithmically enhanced terms are the same. The fixed order results with which one combines the resummed results could be affected through their non logarithmic terms. Unfortunately the commonly used fixed order Monte--Carlo programs do not implement $Z$ exchange and as such we are unable to correct for this effect.

\section{DIS event shape kinematics}
In this section we discuss the main kinematic differences between the observables resummed in this article and those involving the photon axis in the Breit frame \cite{ADS}. 

Consider a hard parton at a small angle $\theta$ to the photon axis ($z$ axis) 
in the Breit frame. This angle is most generally acquired through recoil against emissions entering either hemisphere (current or remnant).
Further consider switching off all secondary 
radiation in the current region. In other words, to start with, just imagine a situation where the emissions are confined to the target hemisphere and the only object in the current hemisphere is the hard parton. 
Note that one imposes a minimum energy requirement in the current region, which guarantees that the current region is not empty and the shape variables are infrared safe. 
Then one obtains for the thrust defined with respect to the photon axis
\begin{equation}
1-T_\gamma = \tau_\gamma = 1-2 \frac{P_z}{Q} \approx 1-\frac{2E}{Q}+\frac{{k_t}^2}{QE} \ ,
\end{equation}
where $k_t$ and $E$ are the parton transverse momentum and energy. 
On the other hand all the variables defined in the previous section vanish ($1-T = \rho=C=0$), unless there are emissions in $\mathcal{H}_C$.
This leads to two effects
\begin{itemize}
\item{If one wants only values of $1-T_\gamma$ near zero, which is the region requiring resummation, one notices that the $k_t$ generated by space-like emissions in the target region has to be small. This is not the case for the other variables here, for example the jet mass still vanishes even though one may have a large $k_t$ (wrt photon axis) single parton in the current hemisphere.
Placing a restriction on the space-like evolution 
means that the scale of the structure function is not $Q^2$ but rather $\tau_\gamma Q^2$. Hence there are $x$ dependent single logarithms that are resummed by DGLAP evolution.
For the other observables the appropriate scale is still $Q^2$.
\item Even though one directly counts only partons (hadrons) in the current hemisphere in the definition of all our observables, the thrust wrt the photon axis and normalised to $Q/2$ is a global observable. It counts soft/collinear secondary gluons in either hemisphere on an equal footing because of recoil effects discussed above\footnote{It has recently 
been realised that the thrust wrt photon axis and normalised to the total current hemisphere energy is a non--global observable in the sense that the emissions in the target and current hemisphere 
are not on the same footing. }.
On the other hand the current jet-mass, the $C$ parameter and the thrust wrt the actual thrust axis are non-global observables sensitive to radiation in only a part of phase space, the current hemisphere.
This necessitates resumming single logs generated by 
coherent emission in the current hemisphere 
from an arbitrary number of soft, energy ordered 
wide-angle gluons in the target hemisphere.}
\end{itemize}
Additionally, since for the non-photon axis variables one has to resum soft/collinear gluons off a hard configuration involving a {\it{wide-angle}} (wrt $\gamma^*$ axis) current jet, we must consider a situation where there is a lone gluon at any angle in the current region which also corresponds to $V=0$. Then one has to resum gluon emissions of this gluonic jet to NLL accuracy.

\section{Resummed results and form factors}
For the jet-mass distribution and event shape cross section one obtains an identical formula to that derived in \cite{dassal}.
For the shape cross section defined as 
\begin{equation}
R(\rho) = \int_0^{\rho} \frac{1}{\sigma}\frac{d\sigma}{d\rho'}d\rho
\def \ee{\end{equation}}'
\end{equation}
we have
\begin{equation}
R(\rho) = (1+\alpha_s C_q^q \otimes q +\alpha_s C_g^q \otimes g)\mathcal{S}(\alpha_s L)\Sigma_q(\alpha_s,L) + \alpha_s C_q^g \otimes q \, \Sigma_g (\alpha_s,L) \ ,
\end{equation}
where $L = \ln\frac{1}{\rho}$.
The form factor $\Sigma_q$, which represents resummation off a hard quark projectile in $\hc$, is required till single logarithmic accuracy 
and can be obtained from \cite{CTTW} 
\begin{equation}
\Sigma_q (\alpha_s,L) = \int_0^{\rho Q^2} J_q (k^2) dk^2 
\end{equation}
and the function $J_q(k^2)$ has been obtained using the coherent branching technique \cite{CTTW}.
Similarly we get the form factor $\Sigma_g$ which represents resummation of a hard gluon jet in $\hc$. 
$\mathcal{S}$ represents the wide-angle soft gluon resummation detailed in \cite{dassal}. 
This function has not been calculated analytically so far but a parametrisation of it in the large $N_c$ limit was obtained by fitting to Monte-Carlo results
\cite{dassal}. Hence one may expect corrections to this piece which are $ \mathcal{O} \left ( \frac{1}{N_c^2} \right )$ (10 \% level).
The various constants $C_a^b$ indicate contributions from incoming quarks or gluons as indicated by the suffix $a$ and convolutions with the appropriate structure functions (pdfs $q$ and $g$). The upper index denotes
 the hard parton in $\hc$ off which one resums soft emissions.

For the thrust and the $C$ parameter resummed results can be derived by noting that they are proportional to the jet mass in the region of interest and one can use the effective relations $ C = 12 \rho = 6(1-T)$ to NLL accuracy.

Before these results can be compared to experimental data from HERA one has to match the results with NLO estimates to get the best possible description over the entire range of values of the variables. This procedure is described extensively in \cite{dassalbroad}.

In order to account for hadronisation effects
the matched distributions are then shifted by $V \to V - a_V {\cal P}$,
applying the concept of power corrections~\cite{dw}.
The coefficients $a_V$ are calculable, while ${\cal P}$ is a universal 
function proportional to $1/Q$ which depends on a non-perturbative
parameter $\an$ as well as on $\as$.
The quantity $a_V {\cal P}$ is exactly the same hadronisation contribution which appears in the corresponding mean values \cite{DasWeb98}.
In fact the analysis of the H1 data~\cite{h1} shows that the $Q$ dependence
of the above defined event shape means can be consistently fitted by a common
parameter $\an \simeq 0.5$ and reasonable values of $\asmz$.
However, fits to event shape spectra using fixed order NLO calculations give in general inconsistent parameters (larger spread) compared to those obtained from the means \cite{martyn}. One reason for the discrepancies may be missing higher order QCD calculations.
The hope is that this situation will improve with the present resummations.
 
\FIGURE{
  \mbox{
    \epsfig{file=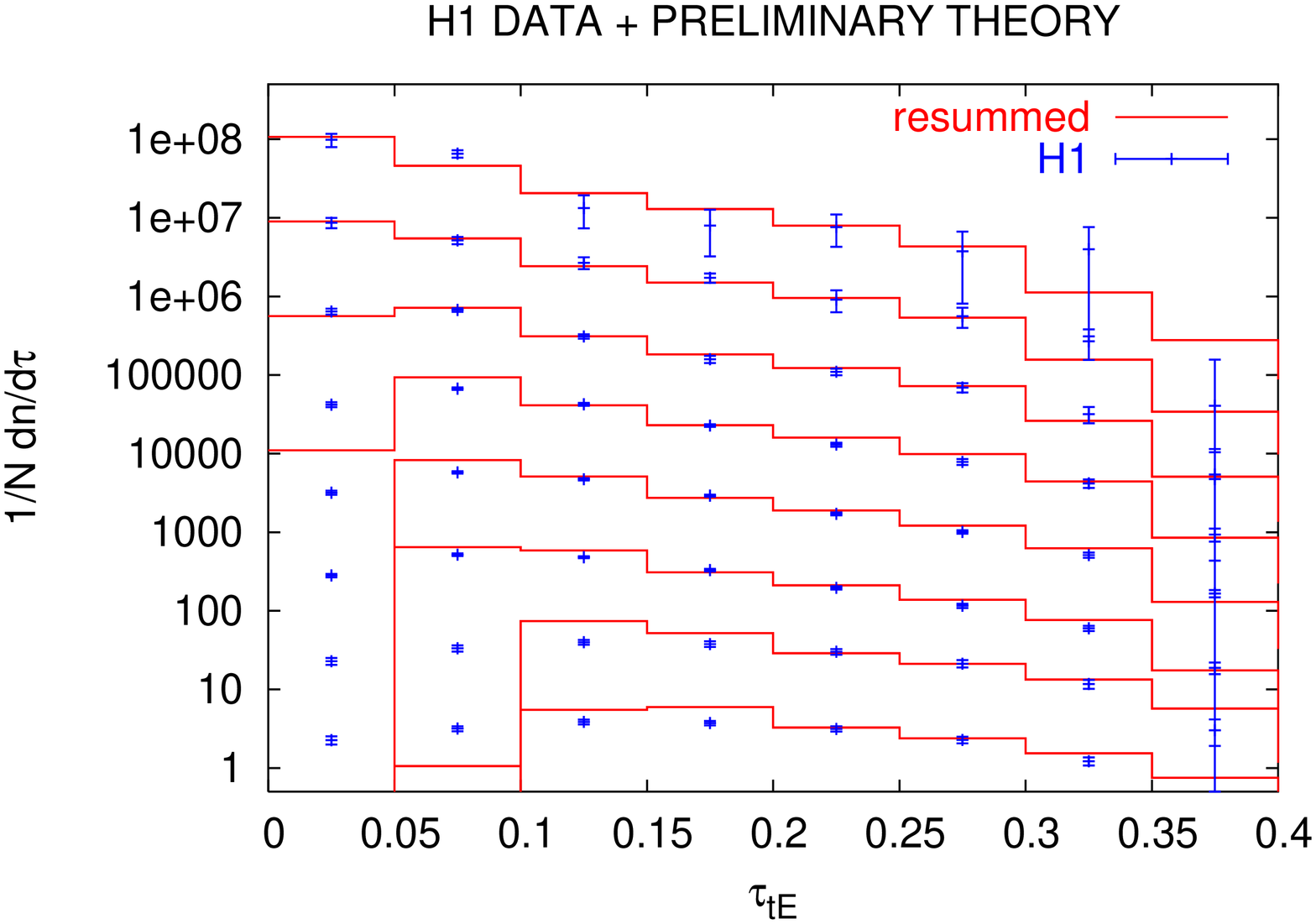,angle=-90,width=.5\textwidth} 
    \epsfig{file=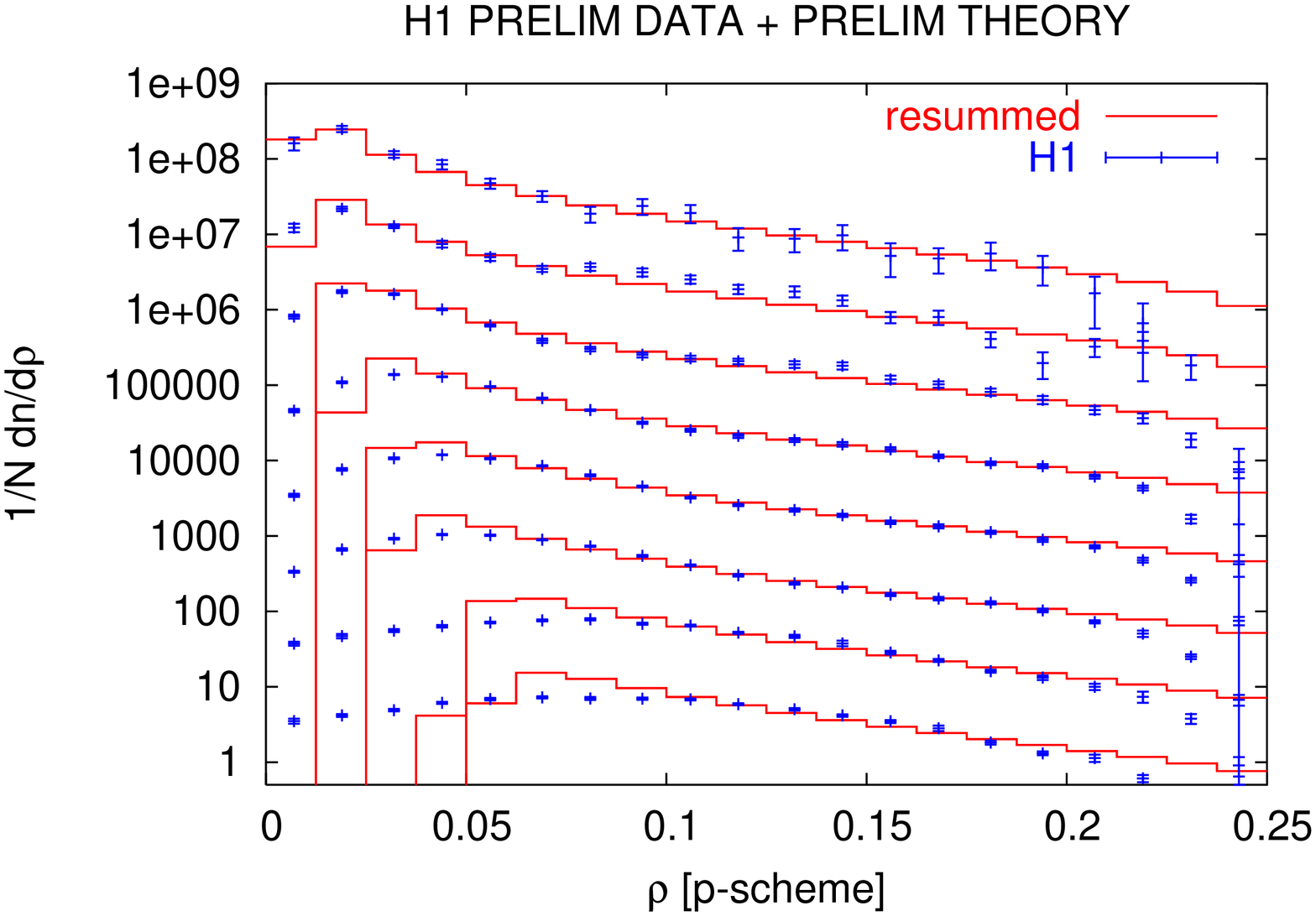,angle=-90,width=.5\textwidth} }
    \caption{Differential distributions of thrust $\tau = 1-T$ (left)
      and jet mass $\rho$ (right).
      H1 data~\cite{h1}
      are compared to resummed QCD calculations including power 
      corrections assuming $\asmz = 0.118$ and $\an = 0.50$.
      The spectra given at $\qmean = 7.5~\GeV$, 8.7 GeV, 15 GeV,
      17.8 GeV, 23.6 GeV, 36.7~GeV, 57.7~GeV and 81.3~GeV 
      (bottom to top) are multiplied by factors of $10^n$ 
      ($n=0,\ldots,7$) }
    \label{fig1} }

Preliminary comparisons with H1 data \cite{h1} on thrust $\tau = 1-T$ and the 
jet mass $\rho$ spectra over a large range of $Q$ values are shown in 
figure~\ref{fig1}.
The resummed distributions are supplemented by power corrections assuming 
$\asmz = 0.118$ and $\an = 0.50$.
The data can be reasonably well described over large parts of the spectrum,
the regions of agreement extend with rising energy $Q$. The improvements of the resummations compared to fixed order calculations are mainly at small values of the event shape variables.
The description deteriorates considerably for lower $Q$ data. This can be understood in terms of the increasing importance of subleading corrections and huge hadronisation effects.

\section*{Summary}
Resummed results are now available for all commonly studied 
event shape spectra in DIS. Preliminary comparisons with data after carrying out matching to NLO and adding power corections look very encouraging. 
A systematic analysis of several DIS event shape distributions including the dependence on scales, matching schemes, etc. is under preparation.

\bigskip\bigskip\noindent {\bf Acknowledgments} \quad
We gratefully acknowledge our collaborators in this work, 
Thomas Kluge, Klaus Rabbertz and Gavin Salam.

\end{document}